\documentclass[11pt,a4paper]{article}
\usepackage[hyperref]{acl2020}
\usepackage{times}
\usepackage{latexsym}
\usepackage{url}

\usepackage{graphicx}
\usepackage{microtype}
\usepackage{subcaption}

\aclfinalcopy

\title{Uncovering Gender Stereotypes in Video Game Character Designs: \\ A Multi-Modal Analysis of Honor of Kings}

\author{
  Bingqing Liu$^{\ast}$\textsuperscript{1},
  Kyrie Zhixuan Zhou$^{\ast}$\textsuperscript{2},
  Danlei Zhu$^{\ast}$\textsuperscript{3},
  Jaihyun Park\textsuperscript{2}\\
  \textsuperscript{1}Dalian Ocean University\\
  liubingqing05@163.com\\
  \textsuperscript{2}University of Illinois at Urbana-Champaign\\
  \{zz78, jaihyun2\}@illinois.edu\\
  \textsuperscript{3}BNU-HKBU United International College\\
  r130201617@mail.uic.edu.cn
}

\date{}

\newcommand\blfootnote[1]{%
  \begingroup
  \renewcommand\thefootnote{}\footnote{#1}%
  \addtocounter{footnote}{-1}%
  \endgroup
}

\begin{document}
\maketitle

\begin{abstract} 
In this paper, we conduct a comprehensive analysis of gender stereotypes in the character design of Honor of Kings, a popular multiplayer online battle arena (MOBA) game in China. We probe gender stereotypes through the lens of role assignments, visual designs, spoken lines, and background stories, combining qualitative analysis and text mining based on the moral foundation theory. Male heroes are commonly designed as masculine fighters with power and female heroes as feminine ``ornaments'' with ideal looks. We contribute with a culture-aware and multi-modal understanding of gender stereotypes in games, leveraging text-, visual-, and role-based evidence. 
\end{abstract}

\section{Introduction}

\blfootnote{$^{\ast}$The first three authors contributed equally to this paper.}

Gender stereotypes, i.e., generalized preconceptions about characteristics or roles of a certain gender, broadly exist in video games, especially competitive games such as League of Legends (LoL) \cite{gao2017gendered}. Honor of Kings\footnote{\url{https://www.honorofkings.com/global-en/}}, the Chinese, mobile counterpart of LoL \cite{cheng2019makes}, was released in 2015. It had over 145 million monthly active users in March 2022 \cite{honor:of:kings} and topped the global mobile game best-selling list with a revenue of 220 million dollars in July 2023 \cite{honor:of:kings:2}. In this game, players can form their teams or randomly match teammates and opponents online. Teammates cooperate to grab resources, kill enemies, and ultimately, destroy the other team's base. Players can collect and play game characters/heroes who are categorized as warriors, assassins, mages, archers, tanks, or supports in the game. 

Gender stereotypes in Honor of Kings have been studied through the lens of the female body \cite{zhang2022female}. Given the prominence of Honor of Kings in the Chinese video game industry, it potentially has wide-ranging impacts on people's conceptions of gender. In this work, we provided a more comprehensive understanding of gender stereotypes in the game character designs through the analyses of hero role assignments, the visual design of heroes and their skins, hero spoken lines, and hero background stories. We conducted a moral foundation analysis \cite{hopp2021extended} on the background stories to understand narratives around human morality, which often sees gendered expectations \cite{zhou2022moral}. We manually analyzed visual designs and hero lines. We also calculated descriptive statistics, i.e., percentages of male and female characters assigned to each role to understand the gender differences in hero role assignments.

We found that female heroes tended to be assigned to more traditionally feminine roles such as mages, while male heroes represented a wider range of roles such as warriors, tanks, and assassins. Female heroes were always designed with idealized looks and body shapes with revealing clothes, while not all male heroes exhibited an idealized appearance. In the spoken lines, male heroes were shaped as being eager to fight and protect, and having supreme power; female heroes were sentimental, objectified, and caring beauties. In the background stories, male heroes were more narrated on authority, while female heroes more on loyalty and sanctity.

Our inspection of gender stereotypes is multi-modal, leveraging text-, visual-, and role-based evidence, and culture-aware, discussing gender roles in ancient Chinese culture. Based on the analysis, we propose future directions to mitigate gender stereotypes in video game character designs.

\section{Related Works} 

The scholarship on games has drawn attention from a wide range of research communities. Some studies revealed the educational benefits of computer games \cite{mayer2019computer} and used games as an instrumental method for student engagement \cite{coller2009video}. Others focused on the collaborative behaviors of players, such as team formation \cite{gomez2019would,kim2017makes}.

There has long been criticism about stereotyping and representation of gender in games. One of the early studies on gender stereotyping in video games was conducted on Super Mario Brothers, where the researcher argued that the narrative of the game could reinforce gender roles as players shared their identities as Mario characters \cite{sherman1997perils}. Another study analyzed ten video games and argued that men were heavily over-represented in games as primary playable characters, power rested on male characters, and female characters remained supportive (e.g., nursing) \cite{friedberg2015gender}. 

At the visual level, Martins et al. found that female characters at low levels of photorealism were larger than the average American woman, while characters at the highest level of photorealism were thinner \cite{martins2009content}. The discrepancy between the real body shape of women and what was portrayed in games could lead to body dissatisfaction in women, attributing the media representation of a thin-ideal body \cite{grabe2008role}. Female character sexualization could lead to self-objectification of female players \cite{fox2015sexualized}, low self-efficacy and self-worth \cite{behm2009effects}, and the acceptance of rape myths \cite{paul2012xbox}.

Researchers have discussed stereotypes in Honor of Kings in terms of the distortion of historical facts. Yao and Chen found hero stories in this game have significantly reconstructed activity processes while largely preserving spatial circumstances, and partly fabricated social relationships among characters, resulting in the distortion of the historical timeline \cite{yao2022reconstructing}. Stereotyping and flattening of the hero images could affect the cultural image of the historical characters \cite{qiu2020stereotyped}.

So far, relatively few studies have focused on gender stereotypes in video games in China \cite{zhang2022female,sun2020gender,chen2023female}. We enrich this literature with the current study. 

\section{Data Collection and Analysis}

\subsection{Data Collection}

As of September 2023, there were 115 heroes in Honor of Kings. Among them, female heroes (N=36) accounted for 31\%, male heroes (N=77) accounted for 67\%, and heroes without a gender presentation (N=2) accounted for 2\%. The skewed gender distribution of heroes already indicated a gender stereotype that men were more suitable and ready for ``wars'' or ``battling'' \cite{hutchings2008making}. 

The \textit{role assignment} and \textit{background story} of heroes were collected from the official website of the game. The \textit{hero lines} were collected from the in-game exhibition of heroes. If the lines of a certain hero were non-verbal, they were excluded from our analysis. The \textit{visual analysis} involved both hero figures and their skins/outfits\footnote{Each hero may have one or more skins; each skin may or may not include a new line.}. Both character lines and background stories are in Chinese. 

\subsection{Data Analysis}

The fast-growing field of Natural Language Processing (NLP) was able in part due to existing datasets and models \cite{park2022raison} as well as metadata in digital archives \cite{dobreski2019remodeling}. To take advantage of existing datasets and models and use them as an analytical lens, we adopted the Chinese Moral Foundation Dictionary (C-MFD) \cite{cheng2023c, wu2019chinese} to analyze the background stories of the heroes. C-MFD can be used for moral intuition detection and analysis in the Chinese language context. The creators of the dictionary drew on the Chinese translation of the English MFD \cite{hopp2021extended} and further fetched related words from an extensive Chinese dictionary based on Chinese moral concepts and word2vec. Categories in C-MFD include care vs. harm, authority vs. subversion, loyalty vs. betrayal, fairness vs. cheating, and sanctity vs. degradation, which are also present in MFD, as well as liberty vs. oppression, waste vs. efficiency, altruism vs. selfishness, diligence vs. laziness, resilience vs. weakness, and modesty vs. arrogance in the Chinese context. 

We calculated moral foundation scores for each hero's \textit{background story} in Chinese and compared the average scores for male and female heroes. One drawback of C-MFD is that it only provides the occurrence frequency for each moral dimension in the text without providing sentiment scores, which prevents us from understanding if the narration of a certain gender leans toward the moral end or the immoral end of a moral dimension.

Since the \textit{hero lines} were less rich in text with short lengths, there was not a sufficient overlap between them and C-MFD. Thus, we analyzed hero lines manually and used the translated lines to showcase our findings. Similarly, we manually analyzed the \textit{visual features} of the heroes and their skins and made cross-gender comparisons. Two authors independently conducted the thematic analysis \cite{braun2012thematic} and regularly discussed to reach a consensus. We used a mind-mapping tool to organize the emerging themes and lines/visual features into a hierarchical structure.

Descriptive statistics were calculated to compare the hero \textit{role assignments} across genders.

\section{Results}
\begin{figure*}[h]
    \centering
    \includegraphics[width=0.6\textwidth]{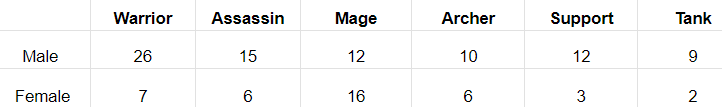}
    \caption{Role assignments for male and female heroes. One hero may have more than one role.}
    \label{role}
\end{figure*}

\begin{figure*}[h]
    \centering
    \includegraphics[width=0.7\textwidth]{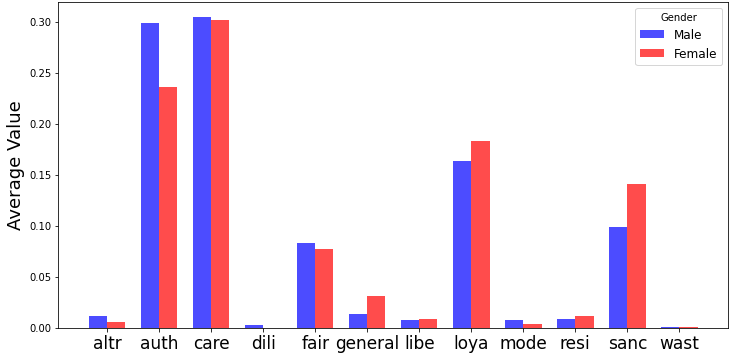}
    \caption{Moral foundation scores for male and female heroes in background stories. From left to right: altruism, authority, care, diligence, fairness, general, liberty, loyalty, modesty, resilience, sanctity, and waste.}
    \label{cmfd}
\end{figure*}

\begin{figure*}[ht]
    \centering
    \includegraphics[width=0.97\textwidth]{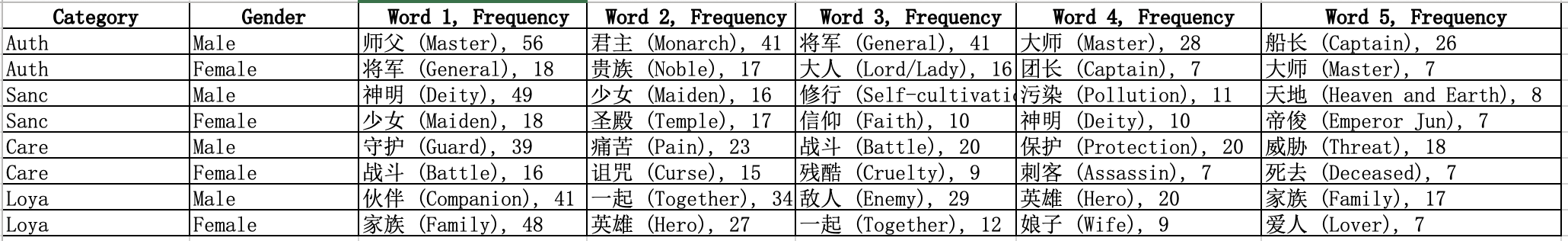}
    \caption{Top 5 moral words for male and female heroes in background stories. We only list top words in moral dimensions with relatively more occurrences.}
    \label{top:5}
\end{figure*}

\subsection{Hero Role Assignment}
There were only 36 female heroes in Honor of Kings compared to 77 male heroes. A closer look at the role assignments for different genders revealed that female heroes were mostly assigned as mages (44\%), who tended to attack and control opponents from a distance. Fewer female heroes were assigned to roles known for hand-to-hand combat, such as warriors, assassins, and tanks. A large portion of male heroes were warriors (34\%), and the remaining male heroes were distributed across other roles. More details are in Figure~\ref{role}.

\subsection{Visual Designs}
Female heroes were designed with standardized physical features that conformed to traditional beauty standards and aesthetic preferences. They were presented as either beautiful or cute, with big breasts, a slim waist, and long legs. Nearly all female heroes had perfect faces that catered to traditional Asian aesthetics, evidenced by a pointy chin, big eyes, a high nasal bridge, a small mouth, and a perfect or even abnormal proportion \cite{zhang2012chinese}. In terms of dressing, we hardly saw female heroes wearing loose clothes. Even if a female hero was a warrior, the designers still intentionally exhibited the curve of her female figure with tight and revealing clothes. Such findings echoed prior studies which found women were portrayed and perceived as sex objects who embodied an idealized image of beauty \cite{dill2007video}.

The height or weight of male heroes varied, but most of them had abdominal muscles on their naked, upper bodies. Although most male heroes were designed to be tall, muscular, or robust to emphasize strength and fighting ability, not all male heroes were traditionally handsome -- some of them were obese, had scars on their faces, or had other stereotypically imperfect characteristics. In general, less focus on the idealized image of beauty was put on male heroes than female heroes. A visual comparison between male and female characters can be seen in Figure~\ref{characters} in the Appendix.

\subsection{Hero Lines}
Most lines, either those of male or female heroes, contained fighting-related elements, possibly due to the battling nature of the game. Yet, we still found differences between genders. The social identities of female heroes were limited to chefs, dancers, or goddesses, and they were often associated with purity and love. Male heroes had more social identities including warriors, princes, musicians, fortune tellers, and so on; they were often associated with conquering, defending, and fighting -- traditionally masculine events.

\subsubsection{Male Heroes}

\textbf{Fighters.} Male heroes' lines were almost always about fighting, war, and violence, e.g., \textit{``We fight for a common tomorrow,'' ``Indomitable soul, inextinguishable fighting spirit, immortal heart,'' ``War soul is not extinguishable,'' ``I'm born for wars.''} Such lines were also spoken with a firm and masculine tone.

\textbf{Protecters.} Male heroes often played the role of a protector for others, including their lovers, homes, and even Earth, e.g., \textit{``Some people want to change the world, while others only want to protect their women,'' ``Saint Seiya will always guard the love and peace of the earth,'' ``In fairy tales, it is said that the prince overcomes thorns to find the imprisoned princess,'' ``Eliminate evil relatives and keep the peace of the world.''}

\textbf{Suprememe Power.} Male heroes' lines indicated the supreme power of men and their self-confidence \cite{meng2023averageyetconfidentmen}, e.g., \textit{``The devil is coming, like my miracle,'' ``Telling you a secret, I'm invincible,'' ``My only flaw is being too perfect.''} 

\subsubsection{Female Heroes}

\textbf{Sentimental.} Many female heroes' lines were about missing their lovers or other sentimental emotions, e.g., \textit{``The east wind sends letters; the flower dynasty is as promised; we see each other every year, and we miss each other every year,'' ``The saying goes that the magpies build a bridge over the cloud, and the destined one will run to you from the other end of the bridge.''} 

\textbf{Serving and Caring.} Female heroes often appeared as caring figures, such as a chef, a housewife, and a waitress. Example lines included \textit{``I'm the one who cooks in the family,''``I add sugar to the memory; guests, please taste it with heart.''} Such lines were uttered in soft, gentle tones by the characters. On the contrary, when a male hero appeared as someone with cooking skills, their lines emphasized the food itself, e.g., \textit{``Only love, justice, and food cannot be disappointed,'' ``Hot and spicy from the depths of the soul.''}

\textbf{Appearance.} Even if a female hero was designed as a warrior, her lines were still about the ``ornamental'' role of women instead of the fighter role, e.g., \textit{``Reap your heart,'' ``Acting as roses,'' ``Yes, I'm tempting you.''} Some lines were explicitly about appearance, e.g., \textit{``I'm so cool and beautiful,'' ``Beautiful girls never look back at explosions.''}

\textbf{Objectified.} Some female heroes' lines exhibited objectification of women, treating women as inferior people or objects, e.g., \textit{``I'm your Christmas present tonight.''} One female hero was designed as a dancer and addressed herself as a ``concubine,'' which was a self-designation in ancient China where women were regarded as possessions of men with a lower social status.

\subsection{Hero Background Stories}
We compared occurrences of moral words between genders and identified notable differences in the authority/subversion, loyalty/betrayal, and sanctity/degradation moral dimensions (see Figure~\ref{cmfd} for a full comparison). Even in moral dimensions with similar occurrence frequency for male and female characters (e.g., care/harm), the gendered narration of heroes was obvious. We further identified the top five common moral words related to these moral dimensions for both genders (Figure~\ref{top:5}). 

\textbf{Authority/Subversion.} Authority/subversion-related moral words were more often seen in the narrative of male heroes than in female heroes. A closer look revealed that male heroes were often narrated with words indicating positions of high authority such as ``master,'' ``monarch,'' ``general,'' and ``captain.'' e.g., \textit{``This is the true face of Master Lu Ban and his genius creation, Lu Ban No. 7!''}  Female heroes were less frequently described as a ``general'' or a ``noble.'' 

\textbf{Loyalty/Betrayal.} Female heroes were more narrated with loyalty/betrayal-related words than male heroes, such as ``family,'' ``wife,'' and ``lover,'' e.g., \textit{``The mission of the family and the responsibilities of the eldest sister fall upon her.''} Male heroes were more linked with such words as ``companion'' and ``enemy,'' again demonstrating their roles as fighters, e.g., \textit{``Every night, he finds himself surrounded by thousands of enemies in his dreams.''}

\textbf{Sanctity/Degradation.} Female heroes were more frequently linked with moral words about sanctity, such as ``maiden'' and ``temple,'' emphasizing the purity and sanctity expectations of women, e.g., \textit{``The maiden feels anger and pain for the unfair treatment.''} Male heroes were also frequently linked with sanctity-related words, such as ``deity''  and ``maiden,'' yet these words suggested their high power, e.g., \textit{``Possess the powerful force of a deity.''}

\textbf{Care/Harm.} Both genders were frequently associated with care/harm-related words, given the battling (harm) nature of this game. Commonly seen in male heroes' stories were ``guard,'' ``pain,'' ``battle,'' ``protection,'' and ``threat,'' showing both tendencies of destruction and protection. Overall, male heroes were narrated as brave and violent, e.g., \textit{``Bajie bravely charged to the forefront of the team.''} On the other hand, female heroes were associated with less powerful words such as ``curse'' and ``deceased,'' e.g.,\textit{``The supreme empress of Chang'an City will never forget her cursed destiny and the dream of an ideal kingdom.''} 

\section{Discussion and Future Work}
By presenting a comprehensive analysis of visual designs, role assignments, spoken lines, and background stories, we uncovered the prevalent gender stereotypes in Honor of Kings.  Our role, text, and visual analyses echoed each other, depicting how the game character design consistently reinforced gender stereotypes. The idealized looks of female characters, combined with the single aesthetic perspective of women in media, i.e., being slim, fair skin, etc. \cite{grabe2008role}, may create pressure and anxiety for women.

In the game, male characters are designed as people in power, fighters, and decision-makers, extending the traditionally perceived role of men in Chinese society \cite{xie2013gender}. Female heroes are designed as feminine ornaments with ideal looks. They tend to play supportive roles (e.g., mage) in battles. Such findings echo prior studies on stereotypes in games \cite{grabe2008role,martins2009content,friedberg2015gender}. Female characters are also stereotypically shaped as emotional and without decisive, competitive, and strong traits.

In traditional/ancient Chinese culture, male superiority and female inferiority were simultaneously emphasized -- men tended to dominate economically and socially, while women were expected to be responsible for childcare and household chores in the family \cite{zhou2022anonymous,zhou2023more,zhou2023public}. The traditional division of gender roles in/outside the family still exists nowadays, though women are increasingly pursuing careers and independence \cite{gui2020leftover}. Such power imbalance and stereotypical gender narration are reflected in Honor of Kings. While prior research criticized the distortion of history in this game \cite{yao2022reconstructing,qiu2020stereotyped}, the gender dynamics are sarcastically true to reality.



With this exploratory uncovering of gender stereotypes in a Chinese video game, we aim to spur more research in this relatively under-investigated cultural context. We suggest several lines of research for future investigation of stereotypes in video game design in specific cultural contexts. First, user studies on how Chinese game players perceive such gender stereotypes are encouraged, as most prior studies were conducted in Western contexts. The persisting, traditional gender roles in East Asian society may lead to either people's desensitization to stereotyping or stronger resistance to it \cite{lee2017multiple}. Second, the video game industry has been long known as a regime of masculine domination \cite{styhre2018masculine,dunlop2007us}. Including more women and gender minorities in the game design and development lifecycle, as well as providing educational interventions to equip designers and developers with gender awareness, are key to mitigating gender stereotypes in games \cite{zhou2023m}. 

\section*{Acknowledgments}
We sincerely thank the anonymous reviewers for their supportive and constructive feedback, which we have leveraged to polish up the paper.

\bibliography{anthology,acl2020}

\begin{thebibliography}{40}
\expandafter\ifx\csname natexlab\endcsname\relax\def\natexlab#1{#1}\fi

\bibitem[{Behm-Morawitz and Mastro(2009)}]{behm2009effects}
Elizabeth Behm-Morawitz and Dana Mastro. 2009.
\newblock The effects of the sexualization of female video game characters on
  gender stereotyping and female self-concept.
\newblock \emph{Sex roles}, 61:808--823.

\bibitem[{Braun and Clarke(2012)}]{braun2012thematic}
Virginia Braun and Victoria Clarke. 2012.
\newblock \emph{Thematic analysis.}
\newblock American Psychological Association.

\bibitem[{Byshonkov(2023)}]{honor:of:kings:2}
Dmitriy Byshonkov. 2023.
\newblock Sensor tower: Top 10 mobile games by revenue and downloads in july
  2023.
\newblock \emph{GameDev Reports}.

\bibitem[{Chen(2023)}]{chen2023female}
Dongna Chen. 2023.
\newblock Female characters’ images in chinese otome game and woman
  stereotype.

\bibitem[{Cheng and Zhang(2023)}]{cheng2023c}
Calvin~Yixiang Cheng and Weiyu Zhang. 2023.
\newblock C-mfd 2.0: Developing a chinese moral foundation dictionary.
\newblock \emph{Computational Communication Research}, 5(2):1.

\bibitem[{Cheng et~al.(2019)Cheng, Yang, Tan, Cheng, Cheng, and
  Zhuang}]{cheng2019makes}
Ziqiang Cheng, Yang Yang, Chenhao Tan, Denny Cheng, Alex Cheng, and Yueting
  Zhuang. 2019.
\newblock What makes a good team? a large-scale study on the effect of team
  composition in honor of kings.
\newblock In \emph{The World Wide Web Conference}, pages 2666--2672.

\bibitem[{Coller and Shernoff(2009)}]{coller2009video}
Brianno~D Coller and David~J Shernoff. 2009.
\newblock Video game-based education in mechanical engineering: A look at
  student engagement.
\newblock \emph{International Journal of Engineering Education}, 25(2):308.

\bibitem[{Dill and Thill(2007)}]{dill2007video}
Karen~E Dill and Kathryn~P Thill. 2007.
\newblock Video game characters and the socialization of gender roles: Young
  people’s perceptions mirror sexist media depictions.
\newblock \emph{Sex roles}, 57(11-12):851--864.

\bibitem[{Dobreski et~al.(2019)Dobreski, Park, Leathers, and
  Qin}]{dobreski2019remodeling}
Brian Dobreski, Jaihyun Park, Alicia Leathers, and Jian Qin. 2019.
\newblock Remodeling archival metadata descriptions for linked archives.
\newblock In \emph{International Conference on Dublin Core and Metadata
  Applications}, pages 1--11.

\bibitem[{Dunlop(2007)}]{dunlop2007us}
Janet~C Dunlop. 2007.
\newblock The us video game industry: Analyzing representation of gender and
  race.
\newblock \emph{International Journal of Technology and Human Interaction
  (IJTHI)}, 3(2):96--109.

\bibitem[{Fox et~al.(2015)Fox, Ralston, Cooper, and Jones}]{fox2015sexualized}
Jesse Fox, Rachel~A Ralston, Cody~K Cooper, and Kaitlyn~A Jones. 2015.
\newblock Sexualized avatars lead to women’s self-objectification and
  acceptance of rape myths.
\newblock \emph{Psychology of Women Quarterly}, 39(3):349--362.

\bibitem[{Friedberg(2015)}]{friedberg2015gender}
Jared Friedberg. 2015.
\newblock Gender games: A content analysis of gender portrayals in modern,
  narrative video games.

\bibitem[{Gao et~al.(2017)Gao, Min, and Shih}]{gao2017gendered}
Gege Gao, Aehong Min, and Patrick~C Shih. 2017.
\newblock Gendered design bias: gender differences of in-game character choice
  and playing style in league of legends.
\newblock In \emph{Proceedings of the 29th Australian Conference on
  Computer-Human Interaction}, pages 307--317.

\bibitem[{G{\'o}mez-Zar{\'a} et~al.(2019)G{\'o}mez-Zar{\'a}, Paras, Twyman,
  Lane, DeChurch, and Contractor}]{gomez2019would}
Diego G{\'o}mez-Zar{\'a}, Matthew Paras, Marlon Twyman, Jacqueline~N Lane,
  Leslie~A DeChurch, and Noshir~S Contractor. 2019.
\newblock Who would you like to work with?
\newblock In \emph{Proceedings of the 2019 CHI conference on human factors in
  computing systems}, pages 1--15.

\bibitem[{Grabe et~al.(2008)Grabe, Ward, and Hyde}]{grabe2008role}
Shelly Grabe, L~Monique Ward, and Janet~Shibley Hyde. 2008.
\newblock The role of the media in body image concerns among women: a
  meta-analysis of experimental and correlational studies.
\newblock \emph{Psychological bulletin}, 134(3):460.

\bibitem[{Gui(2020)}]{gui2020leftover}
Tianhan Gui. 2020.
\newblock “leftover women” or single by choice: Gender role negotiation of
  single professional women in contemporary china.
\newblock \emph{Journal of Family Issues}, 41(11):1956--1978.

\bibitem[{Hopp et~al.(2021)Hopp, Fisher, Cornell, Huskey, and
  Weber}]{hopp2021extended}
Frederic~R Hopp, Jacob~T Fisher, Devin Cornell, Richard Huskey, and Ren{\'e}
  Weber. 2021.
\newblock The extended moral foundations dictionary (emfd): Development and
  applications of a crowd-sourced approach to extracting moral intuitions from
  text.
\newblock \emph{Behavior research methods}, 53:232--246.

\bibitem[{Hutchings(2008)}]{hutchings2008making}
Kimberly Hutchings. 2008.
\newblock Making sense of masculinity and war.
\newblock \emph{Men and Masculinities}, 10(4):389--404.

\bibitem[{Kim et~al.(2017)Kim, Engel, Woolley, Lin, McArthur, and
  Malone}]{kim2017makes}
Young~Ji Kim, David Engel, Anita~Williams Woolley, Jeffrey Yu-Ting Lin, Naomi
  McArthur, and Thomas~W Malone. 2017.
\newblock What makes a strong team? using collective intelligence to predict
  team performance in league of legends.
\newblock In \emph{Proceedings of the 2017 ACM conference on computer supported
  cooperative work and social computing}, pages 2316--2329.

\bibitem[{Lee(2017)}]{lee2017multiple}
Yean-Ju Lee. 2017.
\newblock Multiple dimensions of gender-role attitudes: Diverse patterns among
  four east-asian societies.
\newblock \emph{Family, Work and Wellbeing in Asia}, pages 67--87.

\bibitem[{Martins et~al.(2009)Martins, Williams, Harrison, and
  Ratan}]{martins2009content}
Nicole Martins, Dmitri~C Williams, Kristen Harrison, and Rabindra~A Ratan.
  2009.
\newblock A content analysis of female body imagery in video games.
\newblock \emph{Sex roles}, 61:824--836.

\bibitem[{Mayer(2019)}]{mayer2019computer}
Richard~E Mayer. 2019.
\newblock Computer games in education.
\newblock \emph{Annual review of psychology}, 70:531--549.

\bibitem[{Meng and Literat(2023)}]{meng2023averageyetconfidentmen}
Xingyuan Meng and Ioana Literat. 2023.
\newblock \# averageyetconfidentmen: Chinese stand-up comedy and feminist
  discourse on douyin.
\newblock \emph{Feminist Media Studies}, pages 1--17.

\bibitem[{Park and Jeoung(2022)}]{park2022raison}
Jaihyun Park and Sullam Jeoung. 2022.
\newblock Raison d’{\^e}tre of the benchmark dataset: A survey of current
  practices of benchmark dataset sharing platforms.
\newblock In \emph{Proceedings of NLP Power! The First Workshop on Efficient
  Benchmarking in NLP}, pages 1--10.

\bibitem[{Paul~Stermer and Burkley(2012)}]{paul2012xbox}
S~Paul~Stermer and Melissa Burkley. 2012.
\newblock Xbox or sexbox? an examination of sexualized content in video games.
\newblock \emph{Social and Personality Psychology Compass}, 6(7):525--535.

\bibitem[{Qiu(2020)}]{qiu2020stereotyped}
Zifan Qiu. 2020.
\newblock Stereotyped and flattened: the characteristics and cultural influence
  of hero reconstruction in the game “honor of kings”.
\newblock In \emph{2020 3rd International Conference on Humanities Education
  and Social Sciences (ICHESS 2020)}, pages 131--135. Atlantis Press.

\bibitem[{Sherman(1997)}]{sherman1997perils}
Sharon~R Sherman. 1997.
\newblock Perils of the princess: Gender and genre in video games.
\newblock \emph{Western folklore}, 56(3/4):243--258.

\bibitem[{Styhre et~al.(2018)Styhre, Remneland-Wikhamn, Szczepanska, and
  Ljungberg}]{styhre2018masculine}
Alexander Styhre, Bj{\"o}rn Remneland-Wikhamn, Anna-Maria Szczepanska, and Jan
  Ljungberg. 2018.
\newblock Masculine domination and gender subtexts: The role of female
  professionals in the renewal of the swedish video game industry.
\newblock \emph{Culture and Organization}, 24(3):244--261.

\bibitem[{Sun(2020)}]{sun2020gender}
Jing Sun. 2020.
\newblock Gender in chinese video games.
\newblock \emph{The International Encyclopedia of Gender, Media, and
  Communication}, pages 1--5.

\bibitem[{Wilson(2022)}]{honor:of:kings}
Jason Wilson. 2022.
\newblock Honor of kings is getting even bigger with a global release.
\newblock \emph{Sports Business Journal}.

\bibitem[{Wu et~al.(2019)Wu, Yang, and Zhang}]{wu2019chinese}
S~Wu, C~Yang, and Y~Zhang. 2019.
\newblock The chinese version of moral foundations dictionary: a brief
  introduction and pilot analysis.
\newblock \emph{ChinaXiv}, 10(201911.00002).

\bibitem[{Xie(2013)}]{xie2013gender}
Yue Xie. 2013.
\newblock Gender and family in contemporary china.
\newblock \emph{Population studies center research report}, 13:808.

\bibitem[{Yao and Chen(2022)}]{yao2022reconstructing}
Siyu Yao and Yumin Chen. 2022.
\newblock Reconstructing history and culture in game discourse: A linguistic
  analysis of heroic stories in honor of kings.
\newblock \emph{Games and Culture}, 17(7-8):977--996.

\bibitem[{Zhang(2012)}]{zhang2012chinese}
Meng Zhang. 2012.
\newblock A chinese beauty story: How college women in china negotiate beauty,
  body image, and mass media.
\newblock \emph{Chinese Journal of Communication}, 5(4):437--454.

\bibitem[{Zhang(2022)}]{zhang2022female}
Suoyi Zhang. 2022.
\newblock The female body and experience in chinese multiplayer online battle
  arena games.
\newblock In \emph{2021 International Conference on Social Development and
  Media Communication (SDMC 2021)}, pages 1125--1129. Atlantis Press.

\bibitem[{Zhou et~al.(2023{\natexlab{a}})Zhou, Cao, Yuan, Weissglass,
  Kilhoffer, Sanfilippo, and Tong}]{zhou2023m}
Kyrie~Zhixuan Zhou, Jiaxun Cao, Xiaowen Yuan, Daniel~E Weissglass, Zachary
  Kilhoffer, Madelyn~Rose Sanfilippo, and Xin Tong. 2023{\natexlab{a}}.
\newblock ``i'm not confident in debiasing ai systems since i know too
  little'': Teaching ai creators about gender bias through hands-on tutorials.
\newblock \emph{arXiv preprint arXiv:2309.08121}.

\bibitem[{Zhou and Sanfilippo(2023)}]{zhou2023public}
Kyrie~Zhixuan Zhou and Madelyn~Rose Sanfilippo. 2023.
\newblock Public perceptions of gender bias in large language models: Cases of
  chatgpt and ernie.
\newblock \emph{arXiv preprint arXiv:2309.09120}.

\bibitem[{Zhou et~al.(2023{\natexlab{b}})Zhou, Shen, Zimmer, Xia, and
  Tong}]{zhou2023more}
Zhixuan Zhou, Bohui Shen, Franziska Zimmer, Chuanli Xia, and Xin Tong.
  2023{\natexlab{b}}.
\newblock More than a wife and a mom: A study of mom vlogging practices in
  china.
\newblock In \emph{Companion Publication of the 2023 Conference on Computer
  Supported Cooperative Work and Social Computing}, pages 56--63.

\bibitem[{Zhou et~al.(2022{\natexlab{a}})Zhou, Sun, Pei, Peng, and
  Xiong}]{zhou2022moral}
Zhixuan Zhou, Jiao Sun, Jiaxin Pei, Nanyun Peng, and Jinjun Xiong.
  2022{\natexlab{a}}.
\newblock A moral- and event-centric inspection of gender bias in fairy tales
  at a large scale.
\newblock \emph{arXiv preprint arXiv:2211.14358}.

\bibitem[{Zhou et~al.(2022{\natexlab{b}})Zhou, Wang, and
  Zimmer}]{zhou2022anonymous}
Zhixuan Zhou, Zixin Wang, and Franziska Zimmer. 2022{\natexlab{b}}.
\newblock Anonymous expression in an online community for women in china.
\newblock \emph{arXiv preprint arXiv:2206.07923}.

\end{thebibliography}
\bibliographystyle{acl_natbib}

\appendix

\section{Example Female and Male Characters}
\label{sec:appendix}

\begin{figure}[h]
  \centering
  \begin{subfigure}[b]{0.4\textwidth}
    \includegraphics[width=\textwidth]{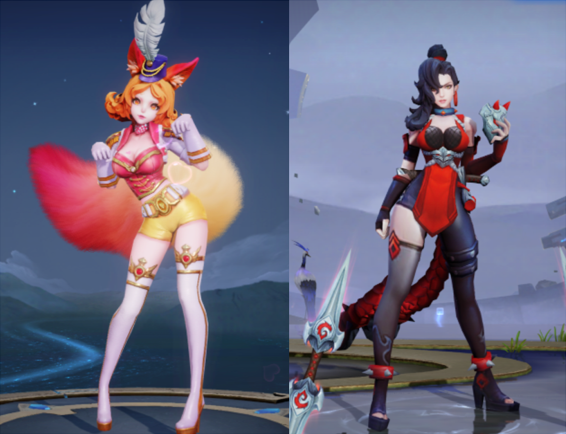}
    \caption{Two female characters. The left is a mage and the right is an assassin.}
    \label{female}
  \end{subfigure}

  \vspace{\baselineskip} 

  \begin{subfigure}[b]{0.4\textwidth}
    \includegraphics[width=\textwidth]{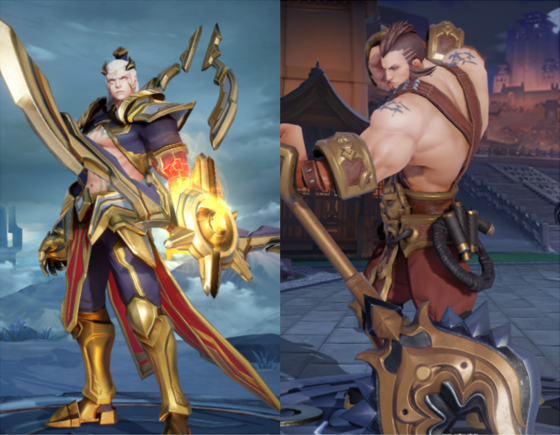}
    \caption{Two male characters. The left is an archer and the right is a tank/warrior.}
    \label{male}
  \end{subfigure}

  \caption{Example female and male characters in Honor of Kings for a visual comparison.}
  \label{characters}
\end{figure}

\end{document}